\documentclass[11pt]{article}

\usepackage[
    letterpaper,
    left=0.85in,
    right=0.85in,
    top=0.85in,
    bottom=1in
]{geometry}

\usepackage{graphicx}
\usepackage{tabularx}
\usepackage{threeparttable}
\usepackage{multirow}
\usepackage{url}
\usepackage{amsmath}
\usepackage{comment}
\usepackage{nameref}

\usepackage{placeins}
\usepackage{authblk}
\usepackage{hyperref}
\usepackage{float}
\usepackage{caption}
\usepackage{subcaption}
\usepackage{amssymb}
\usepackage{amsthm}
\usepackage[english]{babel}
\usepackage{tikz}
\usepackage{lipsum,lmodern}
\usepackage[most]{tcolorbox}
\usepackage{newtxtext,newtxmath}
\usepackage{color}
\usepackage{bm}
\usepackage{booktabs}
\usepackage{placeins}

\title{\textbf{Learning dynamics from online-offline systems of LLM agents}}

\author[1]{Moyi Tian$^*$}
\author[2]{George Mohler}
\author[3]{P. Jeffrey Brantingham}
\author[1]{Nancy Rodr\'iguez}

\affil[1]{Department of Applied Mathematics, University of Colorado, Boulder, CO 80309 United States}
\affil[2]{Department of Computer Science, Boston College, Chestnut Hill, MA 02467 United States}
\affil[3]{Department of Anthropology, University of California, Los Angeles, CA 90095 United States}
\affil[*]{\normalfont Corresponding author: moyi.tian@colorado.edu}

\begin{document}

\maketitle

\begin{abstract}
{Online information is increasingly linked to real-world instability, especially as automated accounts and LLM-based agents help spread and amplify news. In this work, we study how information spreads on networks of Large Language Models (LLMs) using mathematical models. We investigate how different types of offline events, along with the ``personalities’’ assigned to the LLMs, affect the network dynamics of online information spread of the events among the LLMs. We introduce two models: 1) a stochastic agent-based network model and 2) a system of differential equations arising from a mean-field approximation to the agent-based model. We fit these models to simulations of the spread of armed-conflict news on social media, using LLM agents each with one of $32$ personality trait profiles on $k$-regular random networks. Our results indicate that, despite the complexity of the news events, personalities, and LLM behaviors, the overall dynamics of the system are well described by a Susceptible-Infected (SI) type model with two transmission rates.
}
\end{abstract}

\section{Introduction}
Social instability is increasingly connected to a dynamic interplay between online and offline systems. Such instability is causally complex and multiscale in nature, traced to spillover between online and offline domains in both directions. For example, online grievances during the 2013 Kenyan and 2015 Nigerian presidential elections sparked protests and political violence afterward, which then fed back into further online activity \cite{10.1214/20-AOAS1352}. Similarly, gang violence in Chicago has spilled over into negative interactions on Facebook \cite{leverso2025measuring}. These patterns have motivated the use of mathematical and epidemiological contagion models to study social unrest and disorder in the offline setting \cite{bonnasse2018epidemiological,davies2013mathematical,crime_2010} and the spread of information and opinion formation online \cite{rizoiu2018sir,zhao2018extended,fibich2016bass,opinion_dynamics}. More complex online-to-offline spillovers have also been modeled, though in limited ways; see, for example, \cite{peng2021multilayer,leverso2025measuring,10.1214/20-AOAS1352,O2O_Tian}. Extending these modeling approaches, agent-based models that incorporate personality traits have also been used to study information dissemination on networks \cite{Burbach_2019,Burbach_2020}.

While most research has looked at human behavior, automated online activity raises similar concerns. High-volume bot activity is becoming increasingly prominent, with more than half of internet traffic on social networks estimated to be generated by bots \cite{Ng_2025}. As LLM-defined bot behavior becomes more complex and, in some cases, more human-like \cite{AI_collective_behavior}, research is needed on how the parameters associated with LLM bots give rise to pattern formation in large-scale online networks. A recent simulation study found that differences in LLM personalities, set through prompts, affect how fast news spreads and how large the final outbreak becomes \cite{xinyi_2024}. Another experimental study finds that narrative priming, specifically whether agents receive shared or distinct narratives, affects collaboration in network systems of LLM agents during public goods games \cite{Grosmann_2026}. However, these LLM-generated social media dynamics have not been studied from a mathematical modeling perspective, which is the motivation of this work.

\begin{figure}[t!]
    \centering
    \includegraphics[width=\linewidth]{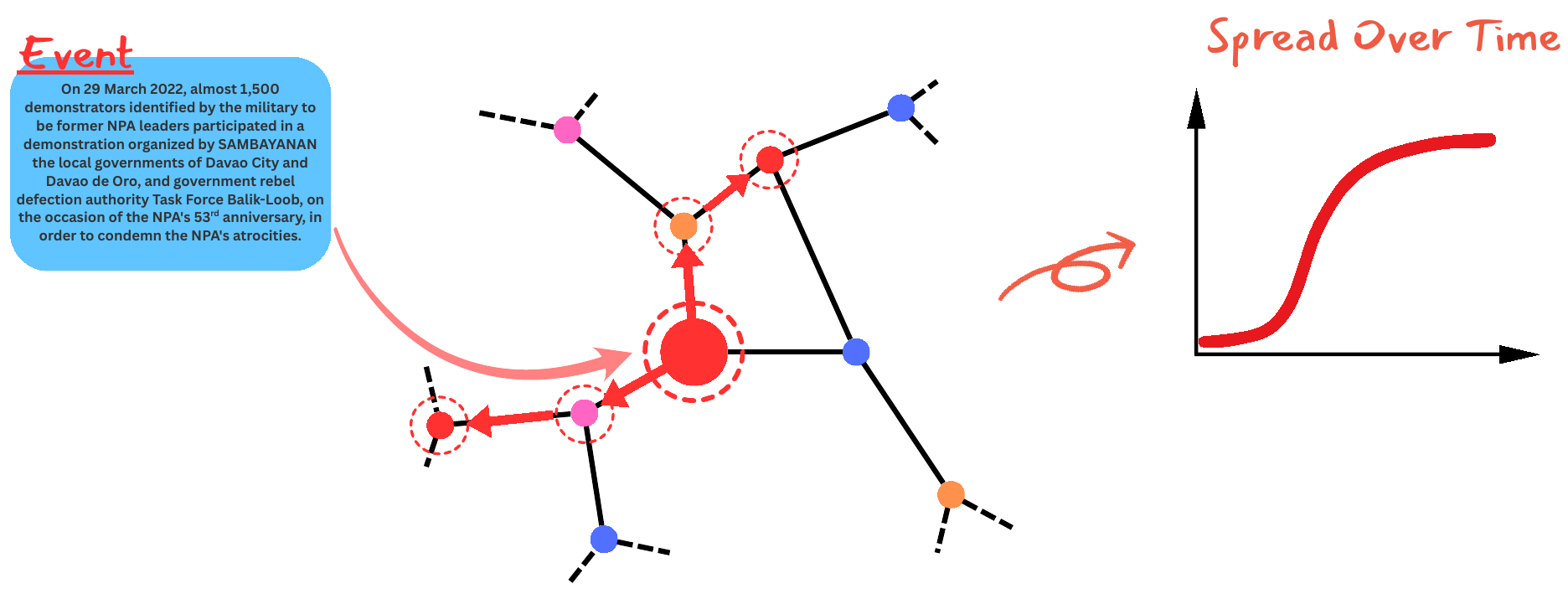}
    \caption{Overview of LLM simulation. LLM agents are connected through a network, each with a pre-assigned ``personality", demonstrated with different colors in the figure. A seed ACLED news event is initially given to one LLM agent, which starts spreading the news. Then, in subsequent steps, the neighbors of the spreaders will randomly decide whether to share it with their neighboring LLMs in the network. The total engagement over time forms a sigmoidal curve (right panel), which we later model using either an agent-based stochastic process or a mean-field approximation.}
    \label{fig:simulation_demo}
\end{figure}

Specifically, we study how AI agents (LLMs) spread information using two models we propose, namely, a stochastic agent-based network model and a mean-field model. We design an experiment in which LLM agents are connected in a network and respond differently to news of protests or violence, leading to distinct global behaviors. An overview of our approach is shown in Figure \ref{fig:simulation_demo}. In each simulation, we seed an initial LLM agent with descriptions of an event from the Armed Conflict Location and Event database (ACLED) \cite{ACLED}. In particular, we chose one event that happened in the Philippines. Each LLM agent randomly decides, based on its prompt-defined personality, whether to share the news with its neighbors, who then similarly make random decisions on passing it or not to their connected neighbors. The aggregated engagement of the LLMs across the network results in a sigmoidal curve over time, which we then link to two mathematical models. The first is a stochastic agent-based network model. The second is a system of differential equations arising from a mean-field approximation of the agent-based model. We study how features of the seed news event and the LLMs’ prompt-defined personalities affect the network dynamics. Our results show that the overall dynamics of engagement can be well described by a Susceptible-Infected (SI) type of model with two groups of agents, each with distinct transmission rates.

{\it Outline:} In Section \ref{sec:method}, we provide details on the LLM simulation, the stochastic agent-based model, and the mean-field approximation. In Section \ref{sec:results}, we perform inference on these models and compare the goodness of fit to the LLM simulation data. In Section \ref{sec:conclusion}, we discuss the implications of our findings and future research directions for mean-field modeling of human-LLM interaction networks.

\section{Methodology}\label{sec:method}
Our method for simulating the spread of information on a network of LLMs with distinct personalities is based on the approach described in \cite{xinyi_2024}. We select news events from the ACLED dataset (details will be provided in Section \ref{news_data}). We categorize events into two main categories: peaceful protests and severe violent incidents that occurred in the Philippines from 2021 to 2024. We describe the $32$ personality types of agents, which arise from binary variables across five traits in Section \ref{agent_profile}. In Section \ref{agent_network_sim}, we discuss the full network simulation, where a seed event is stochastically propagated through the network of LLMs. We present our two mathematical models of the LLM simulation in Section \ref{math_models}. In Section~\ref{agent_model}, we describe an agent-based network model that estimates the likelihood that one agent will pass information to another, using a logistic model based on personality traits and news-related features. In Section~\ref{mean_field_model}, we develop a mean-field approximation for the news-spreading process on networks, which produces a system of differential equations.

\subsection{ACLED event data\label{news_data}}
We seed our network simulation with news events from the ACLED (Armed Conflict Location \& Event Data) database \cite{ACLED}, which collects detailed information on political violence, demonstrations, protests, and other forms of social unrest worldwide. For this study, we focus on events that occurred in the Philippines between February 24, 2021, and February 27, 2024. We provide some sample descriptions of these events in Section \ref{agent_profile}.

\begin{figure}[!htbp]
    \centering
    \includegraphics[width=\linewidth]{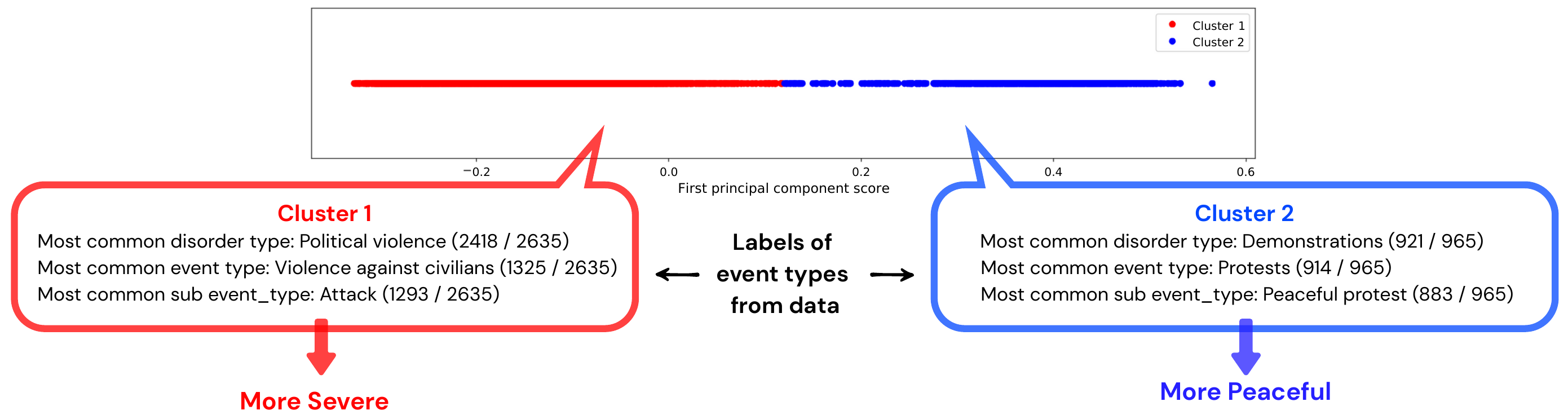}
    \caption{The distribution of events as clustered into $2$ groups by the first principal component of the text embeddings, suggested by possessing the highest Silhouette score. The clusters have distinct levels of severity: cluster $1$ contains more severe events, with the most common event-type labels being ``political violence", ``violence against civilians", and ``attack". In contrast, events in cluster $2$ appear to be more peaceful, with the most common labels being ``demonstrations", ``protests", and ``peaceful protest".}
    \label{fig:event_cluster}
\end{figure}

The transmission of ACLED news events by LLM agents results from an interaction between the news text and the agent's personality defined through the prompt. To parameterize this interaction, we first convert each event's text description to an embedding vector using the \textit{text-embedding-3-small} model from \textit{OpenAI} \cite{openai_text_embedding_3_small}. Next, we apply principal component analysis to the text embeddings. We find that a single principal component with two clusters yields the highest Silhouette score \cite{shahapure2020cluster}, which ranges from $-1$ to $1$ with higher values indicating better clustering quality. This motivates us to group the events into two broad clusters based on the first principal component of the text embeddings. The ACLED event data include labels for event classifications at three levels of granularity: disorder type, event type, and sub-event type. Disorder type groups events into broad domains of political contention, such as ``political violence", ``demonstrations", and ``strategic developments". Event type further groups events into primary forms of behavior observed during the event, such as ``battles", ``violence against civilians", and ``protests". Sub-event type refers to the tactical or operational modality, such as ``armed clash", ``arrests", and ``peaceful protest". By inspecting the labels of the events in each cluster, we find that the events are mainly divided by severity: the first cluster comprises more severe attacks and violence, whereas the second cluster consists largely of peaceful demonstrations and protests. In cluster $1$, the most common categories, one in each of the three event type levels, are ``political violence", ``violence against civilians", and ``attack". In cluster $2$, the most common categories are ``demonstrations", ``protests", and ``peaceful protest", respectively. Figure \ref{fig:event_cluster} illustrates the event clustering along the first principal component of the text embeddings. This binary cluster variable is then used as a feature within the mathematical models detailed in Section \ref{math_models}.

\subsection{Prompt-defined agent profiles \label{agent_profile}}
\begin{figure}[!htbp]
    \centering
    \begin{subfigure}[t]{0.32\textwidth}
        \centering
        \includegraphics[width=\linewidth]{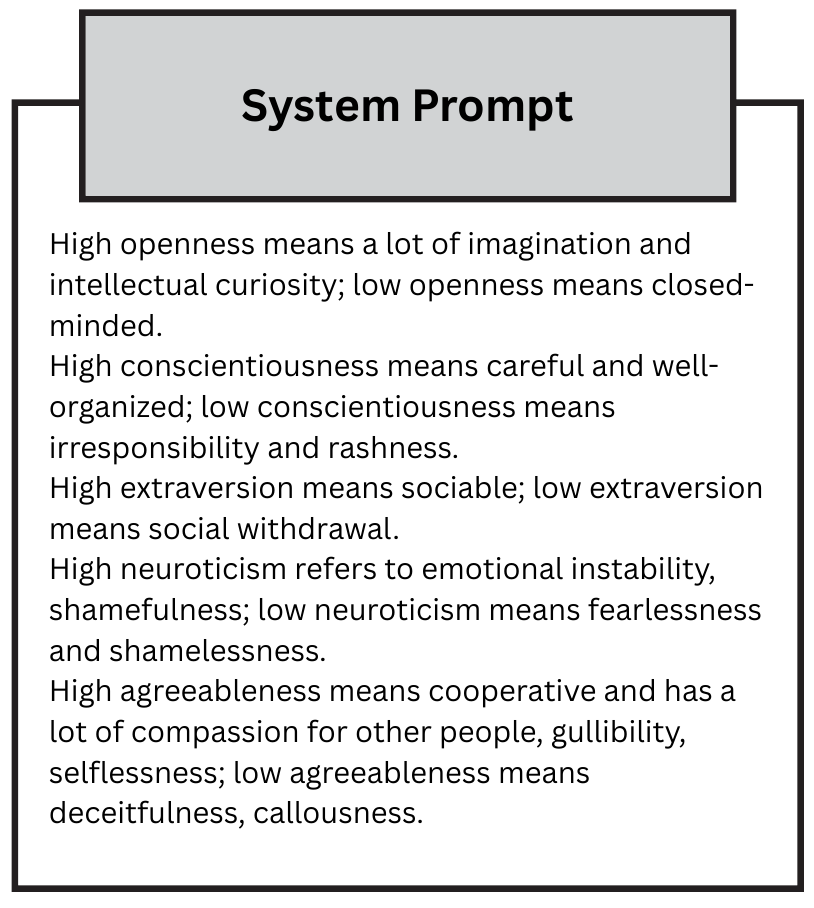}
        \caption{System prompt}
        \label{subfig:system_prompt}
    \end{subfigure}
    \hfill
    \begin{subfigure}[t]{0.32\textwidth}
        \centering
        \includegraphics[width=\linewidth]{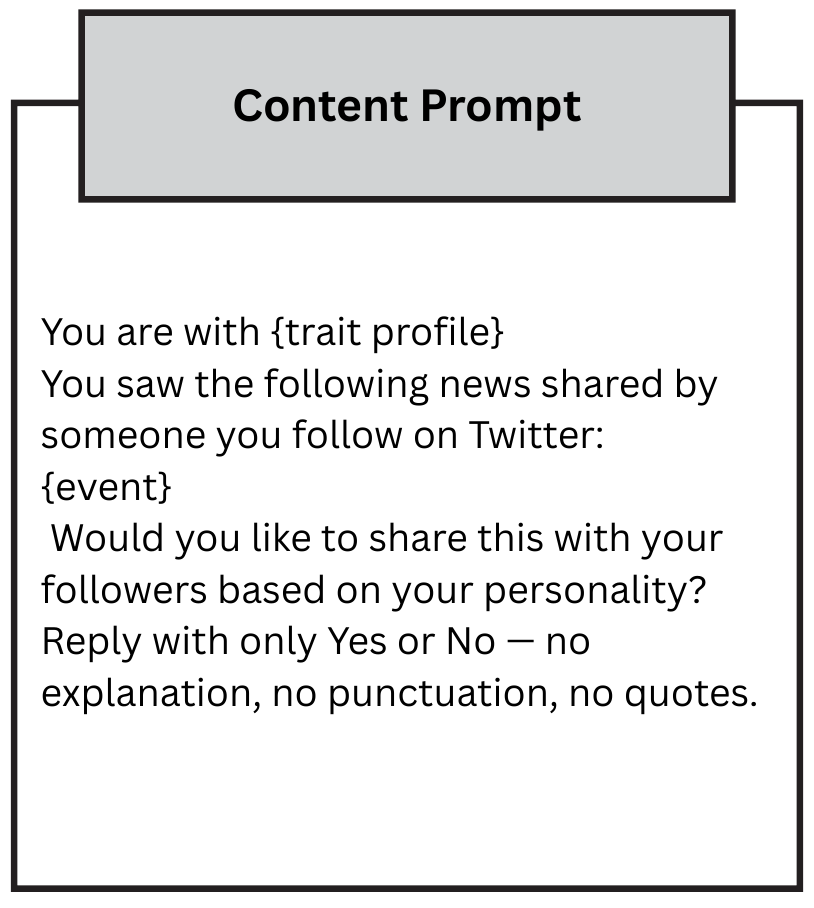}
        \caption{Content prompt}
        \label{subfig:content_prompt}
    \end{subfigure}
    \hfill
    \begin{subfigure}[t]{0.32\textwidth}
        \centering
        \includegraphics[width=\linewidth]{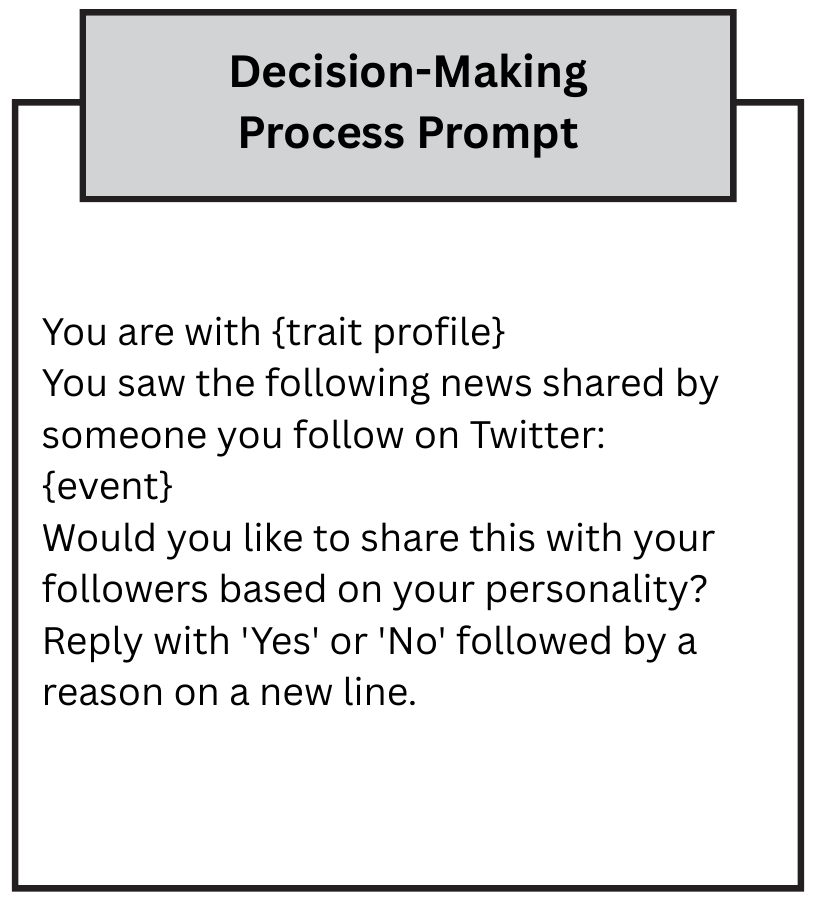}
        \caption{Decision-making process prompt}
        \label{subfig:decision_making_prompt}
    \end{subfigure}
    \caption{Prompts to the LLMs that define personality behaviors and ask about their responses based on events. (a) The system-level prompt that defines the expected behaviors of the LLM's Big Five personality traits. (b) The agent-level prompt that gives the news text and asks the LLM to decide whether to share it with neighbors based on its personality. (c) The post-simulation prompt that asks the LLM to explain its decision. }
    \label{fig:prompts}
\end{figure}

The Big Five Model is a well-established framework in differential psychology \cite{Big5_taxonomy_1963, Big5_analysis_1981, Big5_study_1990, Big5_book_2000} that describes personality across five key dimensions: openness (willingness to try new experiences), conscientiousness (organization and self-discipline), extraversion (sociability and assertiveness), agreeableness (cooperativeness and compassion), and neuroticism (emotional stability and tendency toward negative emotions). Following the work in \cite{xinyi_2024}, we assign each agent a binary trait profile across these five dimensions of the Big Five Model. Each dimension is assigned a binary variable indicating whether it is low or high. The profile assignment is random and independent across agents and simulations. In Figure \ref{fig:prompts}, we display the LLM prompts defining the personality traits, along with the prompts that share the news text with the LLM agent and ask the agent whether or not it wants to share the news with neighboring agents in the network, and provide the reasoning.


\begin{figure}[!t]
    \centering
    \begin{subfigure}[t]{0.99\textwidth}
        \centering
        \includegraphics[width=\linewidth]{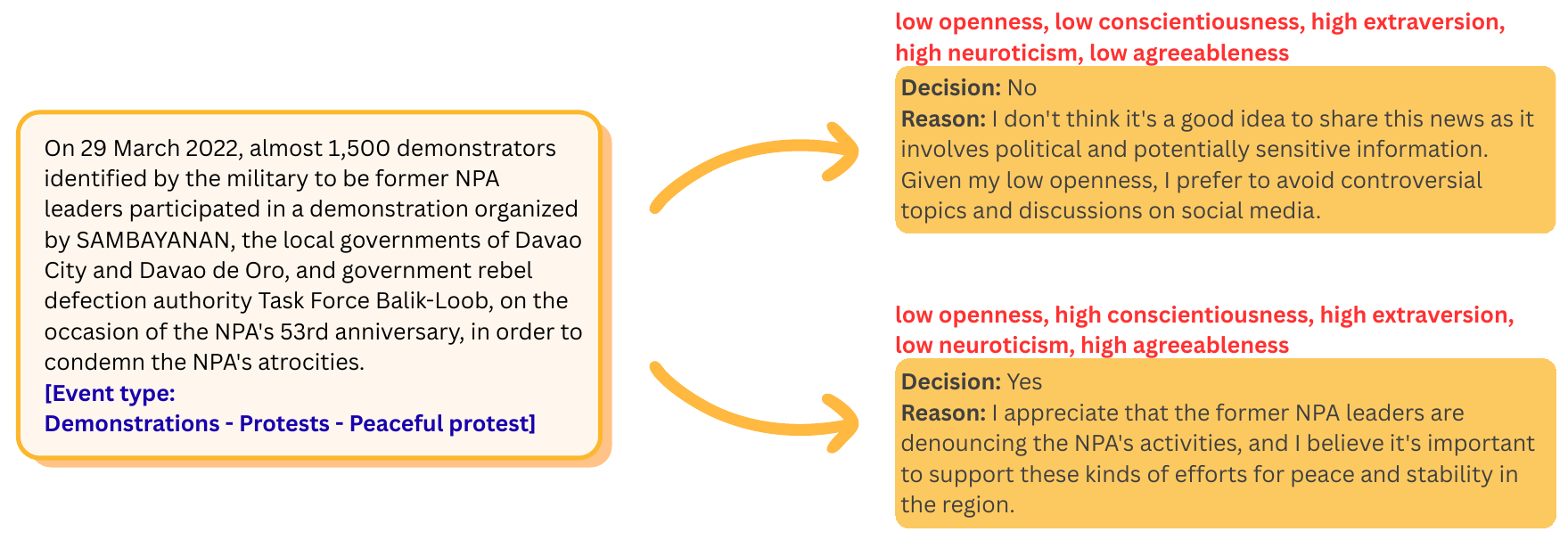}
        \caption{Peaceful event}
        \label{subfig:decision_making_eg1}
    \end{subfigure}
    
    \vspace{1em}
    
    \begin{subfigure}[t]{0.99\textwidth}
        \centering
        \includegraphics[width=\linewidth]{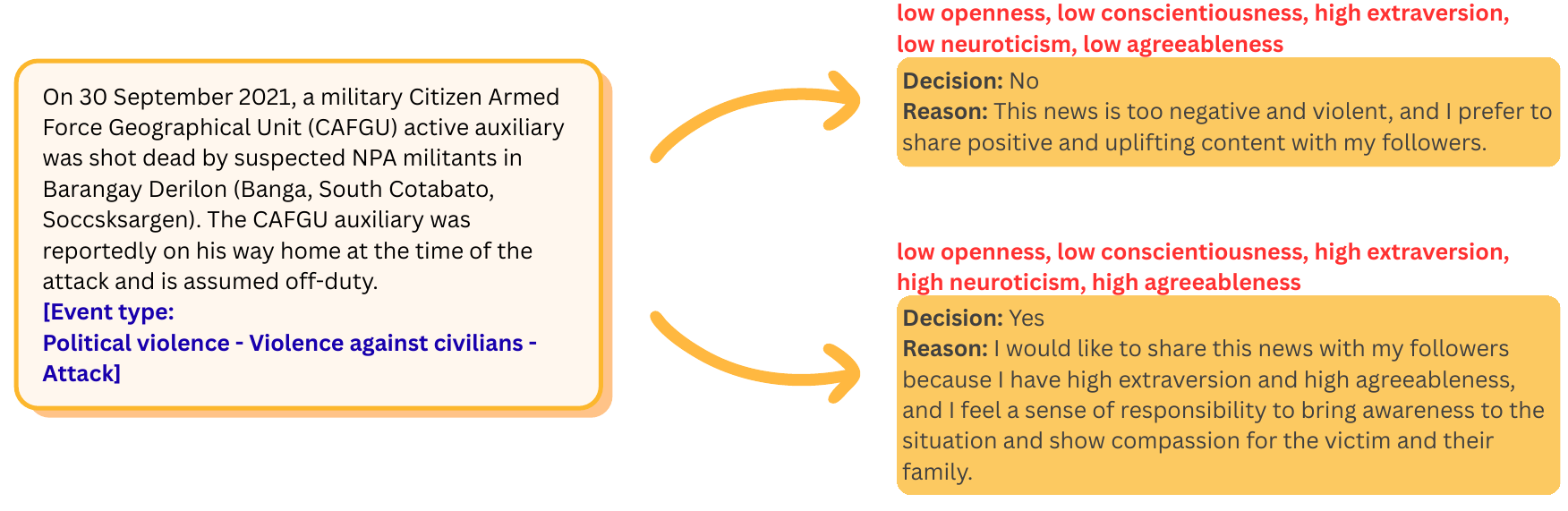}
        \caption{Severe event}
        \label{subfig:decision_making_eg2}
    \end{subfigure}
    
    \caption{The left column presents ACLED event news descriptions for an example peaceful protest (top) and violent attack (bottom). News is shared with LLM agents via a prompt, and a decision is then made to share or not share the news with neighbors in the network (depending on the personality profile). On the right, examples of agents with different personalities are shown with their decisions and rationales.}
    \label{fig:decision_making_egs}
\end{figure}

In post-simulation experiments, we ask LLM agents to explain why they chose to share or not share the news using the prompt shown in Figure~\ref{subfig:decision_making_prompt}. For example, an agent with low openness might choose to avoid sharing controversial topics. Figure \ref{fig:decision_making_egs} shows some examples of the prompt and decision process. ACLED events, whether peaceful or violent, are passed to each agent when their neighbors in the network share the news. Each agent then decides whether to further share it with their corresponding neighbors based on their personality and the news content, and provides a reason for that decision.


\subsection{Interacting agent network simulation \label{agent_network_sim}}
To distinguish key spreading patterns due to event features or inclinations from personalities from potential randomness arising from network connectivity and bias in the distribution of personality profiles, in our setup, we maintain some level of randomness to reflect the noisy nature while keeping certain settings under control to make results more interpretable. Specifically, we generate $k$-regular random networks with $N=128$ agents and degree $k=5$. We use \textit{gpt-3.5-turbo-1106} from \textit{OpenAI} \cite{openai_gpt35_turbo_1106} to model agents. To obtain randomness in responses that reflect creativity and noise in human behavior, we set the temperature to $(0.7, 1.3)$, a moderate range that preserves the key behavioral tendencies associated with the assigned personality while introducing a natural degree of noise in decision-making. Agents are assigned one of $32$ possible trait profiles, since there are five personality traits and each is assigned a binary variable independently. We distribute the $32$ personality profiles uniformly at random across agents on the network; in a network of $128$ agents, each profile appears $4$ times.

At the start of each simulation, we randomly select one agent and run through the seed prompt, asking whether this agent is willing to share the event with its online followers (i.e., its neighbors on the network). If not, we continue to select a new agent until the response is yes. We label this initial agent as ``engaged" and every other agent as ``unengaged". In each subsequent iteration of the simulation, we ask every ``unengaged" agent, as long as it has some ``engaged" neighbors, whether it wants to further share the event with its neighbors, provided there remain neighbors who have yet to receive the news. Each agent can be prompted multiple times, as the number of prompts at each iteration corresponds to the number of ``engaged" neighbors they have. Thus if one ``unengaged" agent has more ``engaged" neighbors, it will be prompted more times, increasing the probability of spreading the news. If the agent decides to share the event, its label is switched to ``engaged" from ``unengaged". We assume that once an agent becomes ``engaged", it never switches back to ``unengaged". Therefore, the engagement ratio is monotonically increasing over time. 

Each network simulation is run for $T=15$ iterations to capture key dynamics while keeping runtime reasonable. We run multiple realizations using $10$ seed events ($5$ from each event cluster). For each seed event, we repeat the entire simulation process $5$ times, including the random generation of the network and the random assignment of traits.

\subsection{Mathematical models\label{math_models}}
The LLM network simulation described in the previous section is characterized by a contagion process in which, at each iteration, each node $i$ in the network is associated with a probability $p_i$ of sharing news. This probability is conditioned on whether neighbors themselves share news, and depends on the personality of the LLM (defined through the prompt), the content of the news, and the LLM model parameters (along with the temperature). In Section~\ref{agent_model}, we present a logistic regression for modeling $p_i$ as a function of covariates associated with the news text and the personality trait vector. Then, in Section~\ref{mean_field_model}, we present a mean-field approximation to the agent-based model.

\subsubsection{Agent-based stochastic network model\label{agent_model}}
We propose a logistic model that describes the probability of an ``unengaged" node becoming ``engaged" upon exposure to an engaged neighbor based on personality trait profiles and event severity.  Let $y \in \{0,1\}$ denote the binary response indicating news being shared ($0 =$ ``no'', $1 =$ ``yes''), $\mathbf{t} \in \{0,1\}^5 = \{ (b_1,b_2,b_3,b_4,b_5) \mid b_i \in \{0,1\}, i=1,2,3,4,5 \}$, the binary 5-d trait vector of the node ($0 =$ low, $1 =$ high), $s \in \{0,1\}$ the event severity label ($0 =$ peaceful, $1 =$ severe), $\beta_{\mathbf{t}}$ the weight specific to trait profile $\mathbf{t}$, and $\beta_s$ the coefficient for event severity. The probability of engagement given traits $\mathbf{t}$ and event severity $s$ is modeled as
\begin{align} \label{model:logistic_32}
    \Pr(y = 1 \mid \mathbf{t}, s) = \sigma\left(\sum_{\mathbf{b} \in \{0,1\}^5}\mathbf{1}_{ \{\mathbf{t} = \mathbf{b} \}} \cdot \beta_{\mathbf{b}} + \beta_s \cdot s\right),
\end{align}
where $\sigma(z) = \frac{1}{1 + e^{-z}}$ is the logistic sigmoid function.

Given $N$ datapoints $\{(\mathbf{t}_i, s_i, y_i)\}_{i=1}^N$, we estimate the parameters by minimizing the negative log-likelihood
\[
\mathscr{L} = - \sum_{i=1}^N \left[ y_i \log p_i + (1 - y_i) \log(1 - p_i) \right],
\]
where $p_i = \Pr(y_i \mid \mathbf{t}_i, s_i)$ as determined by Equation \eqref{model:logistic_32}.  The parameters to be inferred are $\{\beta_{\mathbf{t}}\} \in \mathbb{R}^{32}$ (one coefficient for each trait profile), and $\beta_s \in \mathbb{R}$ (severity effect). We perform maximum likelihood estimation to infer these parameters using the training dataset from the network simulation.

\begin{figure}[!t]
    \centering
    \includegraphics[width=\linewidth]{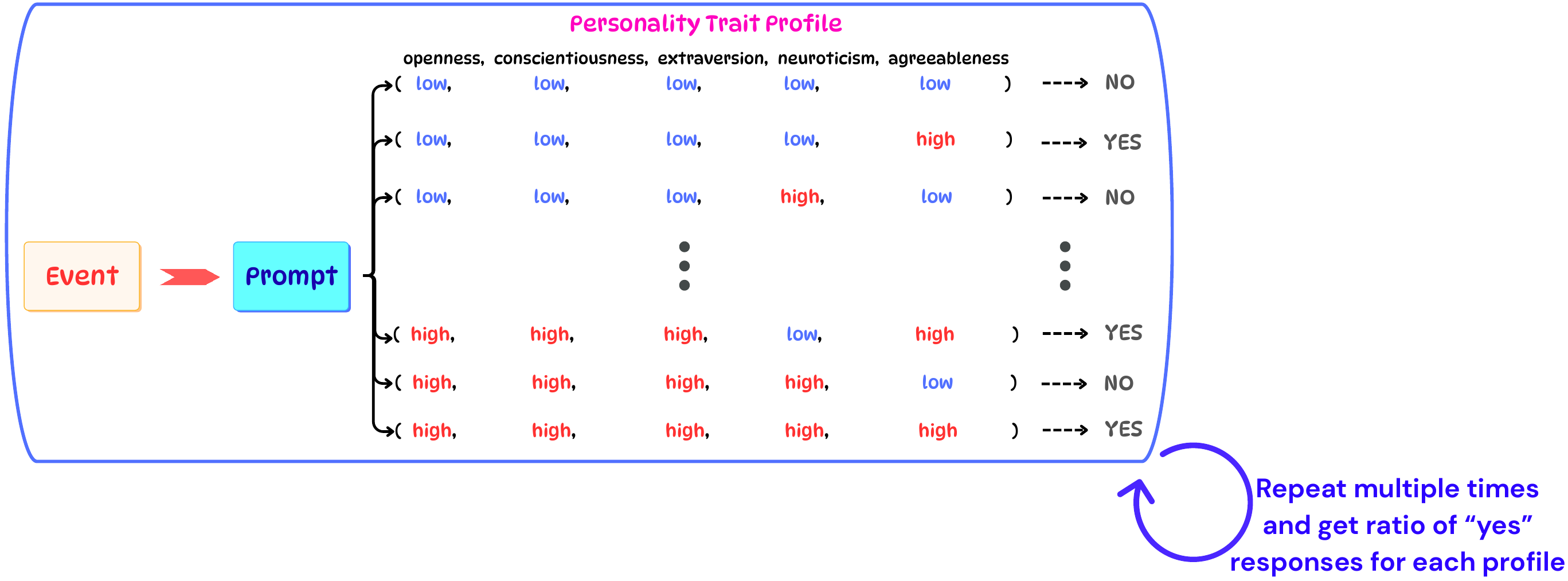}
    \caption{Demonstration of the process obtaining the response ratios for each personality profile for each given event to be used to later classify traits based on spreading behaviors. For a given event, agents with different personalities will each be prompted multiple times to retrieve a ``yes" response ratio related to that event and that personality.}
    \label{fig:response_ratio_process}
\end{figure}

To group the trait profiles based on spreading behaviors, we additionally run another set of simulations in which we apply the decision prompt on $25$ events from each of the two event clusters. Then, for each event, we run the prompt for each of the $32$ personality profiles $10$ times. Figure \ref{fig:response_ratio_process} illustrates the process for obtaining response ratios of willingness to spread for different events across the $32$ personality profiles.

We explore two logistic models: a full logistic regression with 32 personality profiles, and a coarse-grained logistic regression with personality clusters. For the latter, we use the \textit{Tukey HSD test} \cite{abdi2010tukey} to identify clusters of personality profiles based on their response behaviors, aggregated from data obtained in the aforementioned process, that are statistically significant. The test yields two personality groups, where the average sharing rate is $0.0533$ for group 1 and $0.8154$ for group 2.  We include results for the full logistic regression model in Section \ref{sec:results} and results for the coarse-grained two-group logistic model in Appendix \ref{app:logistic_trait_group_based}.

\subsubsection{Mean-field differential equation model\label{mean_field_model}}
Next, we study a mean-field approximation of the news-spreading process on networks, following a Susceptible-Infected (SI) type model in epidemiology. Mean-field approximations of dynamics on networks allow us to draw insights from a simpler model while still capturing the global behavior of the system \cite{eames2002modeling, Epidemics_Networks}. 

Let $U$ denote the population ratio in the ``unengaged" state, and let $E$ denote that in the ``engaged" state. Without splitting agents into groups, we obtain a mean-field approximation model for the population-level $U$--$E$ system on a $k$-regular graph by assuming that the engaged agents are distributed randomly across the network, and so, on average, the number of unengaged neighbors of an engaged agent will be $kU$ \cite{O2O_Tian}. We thus arrive at the governing differential equations
\begin{align*} 
\frac{dU}{dt} &= -k \tau U E, \quad 
\frac{dE}{dt} = k \tau U E,
\end{align*}
where $\tau$ is the edge-level engagement rate, $k$ is the node degree, and $U(t)$ and $E(t)$ denote the fractions of unengaged and engaged individuals at time $t$, respectively.

We extend this system to incorporate the trait structure. Since we distinguish individuals by the two coarse-grained trait groups as discussed in the previous section, we consider the system
\begin{align} 
\begin{cases} \label{model:MF}
\dfrac{dU_1}{dt} = - k \tau_1 U_1 E, \\[1ex]
\dfrac{dU_2}{dt} = - k \tau_2 U_2 E, \\[1ex]
\displaystyle\dfrac{dE}{dt} = \sum_{i = 1,2} k \tau_i U_i E,
\end{cases}
\end{align}
where $U_i$ is the ratio of individuals in the ``unengaged" state with trait in group $i$ over the full population, and $\tau_i$ is the edge-based spreading rate for trait group $i$. Specifically, $\tau_i$ governs the rate at which an unengaged individual with profile in group $i$ becomes engaged through connection with an engaged neighbor.

\begin{figure}[!htbp]
    \centering
    \includegraphics[width=0.55\linewidth]{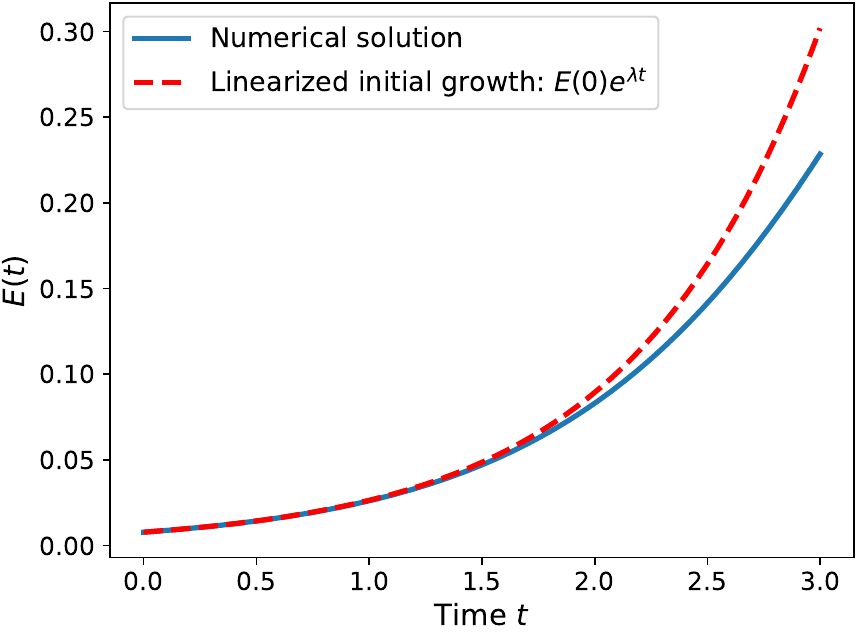}
    \caption{Numerical solution of the proportion of engagement by system \eqref{model:MF} and the growth rate determined by the eigenvalue, $\lambda$, associated with the unstable direction in $E$ at the spread-free equilibrium. During the initial period of time, $\lambda$ can approximate the exponential growth rate of the dynamics when the system has a small initial $E(0)$.}
    \label{fig:InitGrowE}
\end{figure}

For system \eqref{model:MF}, we can explicitly compute that there are two equilibrium solutions, the spread-free state $(U_1, U_2, E) = (a, 1-a, 0)$ for $a \in [0,1]$ and the fully engaged state $(U_1, U_2, E) = (0,0,1)$. By applying the mass conservation constraint $U_1 + U_2 + E = 1$, we can reduce the system to 2D and more easily analyze the stability of these equilibrium solutions. The spread-free state is unstable, while the fully engaged state is stable, meaning that, in the long term, we expect every agent in the system to be engaged. Figure \ref{fig:InitGrowE} shows that the eigenvalue, $\lambda$, associated with the unstable direction in $E$ for the spread-free state can be used to approximate the growth rate of the engagement during the initial period of time when the system kicks off with a small number of spreaders. 

Notice that in this model, we use least-squares fitting to estimate the only two parameters, $\tau_1$ and $\tau_2$. In particular, we minimize the square error between the differential equation engagement curve and the observed engagement curve averaged across network simulations. This contrasts with the full agent-based model \eqref{model:logistic_32}, where we estimate a logistic regression via MLE and infer $33$ parameters.

\section{Results}\label{sec:results}
In this section, we present the results of the logistic model and the mean-field model.  

\subsection{Agent-based model results}
\begin{figure}[!htbp]
    \centering
    \includegraphics[width=\linewidth]{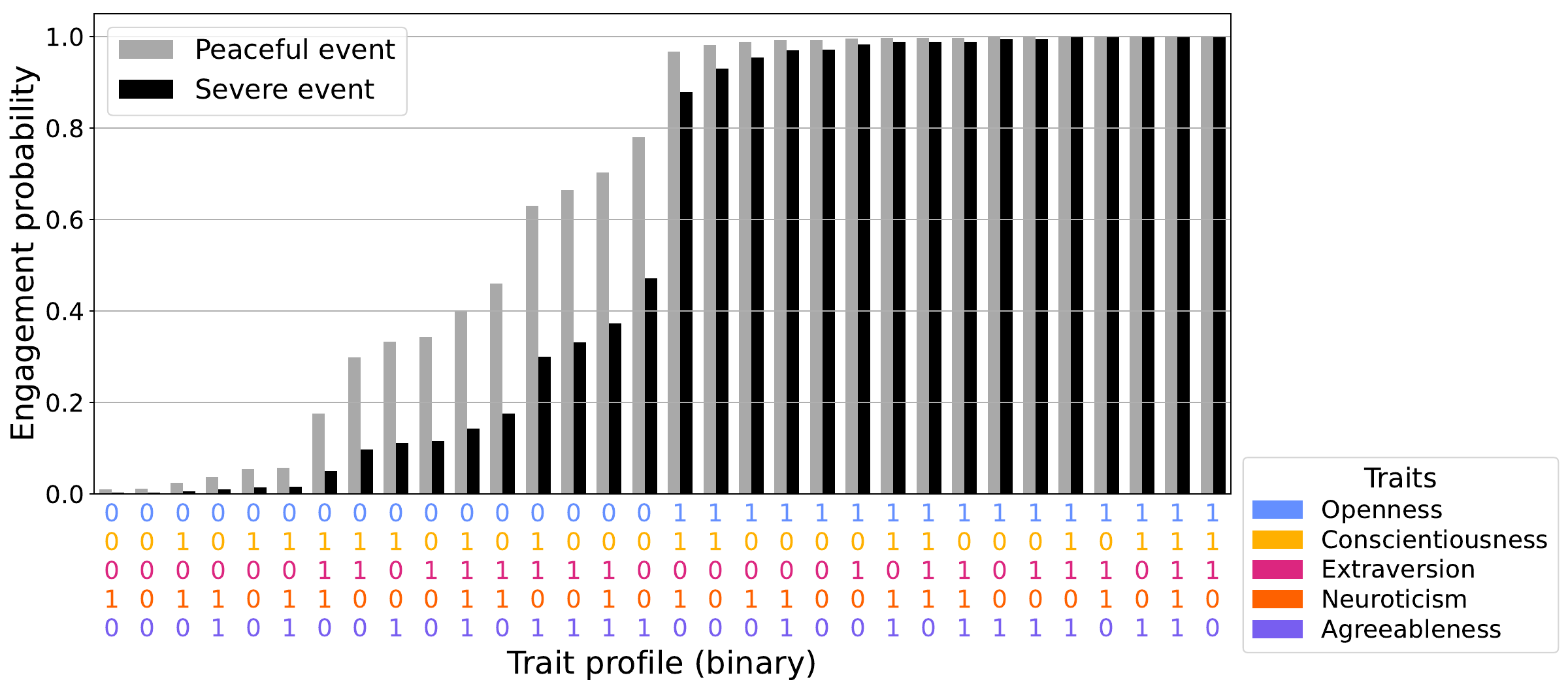}
    \caption{Engagement probability vs trait profile disaggregated by event type, based on inferred model \eqref{model:logistic_32} using training data. In the binary string, $0$ means the corresponding trait dimension is ``low", while a $1$ means ``high."}
    \label{fig:engage_prob}
\end{figure}

We begin by analyzing the logistic model fit to the LLM simulation data. In Figure \ref{fig:engage_prob}, we plot the probability of engagement, which is the probability of spreading the news to neighbors, for each personality trait and event type as computed based on results of the inferred model \eqref{model:logistic_32}. Here, we observe that LLM agents are more likely to share information concerning peaceful events, compared to events that are more severe/violent. We also observe that the probability of sharing news varies significantly across different personality traits.

Figure \ref{fig:engage_prob} also suggests that some dimensions of personality traits play a significant role in determining the spreading behavior of agents. For example, we see that a sharp, non-linear jump in the engagement probability for severe events is observed when openness first changes from low to high. 

To better understand how each of the five Big Five dimensions influences the personality profiles in \eqref{model:logistic_32}, we also fit a simpler logistic model using the same training data. Instead of estimating all 32 personality interaction weights, this complementary model is linear with respect to the five personality traits, and therefore has 5 personality features, along with one parameter for event severity and one parameter for the intercept.

Table \ref{tab:p-vals} shows the inferred weights and their results from statistical significance testing for this simpler logistic model. Same as our observation of Figure \ref{fig:engage_prob}, the p-values in the table indicate that openness has the largest coefficient and smallest p-value, indicating that it is the strongest determining factor of the transmission rate, followed by extraversion among the five personality trait dimensions.

\begin{table}[!htb]
\centering
\caption{Logistic regression coefficients and significance tests for personality dimensions, event severity, and intercept.}
\label{tab:p-vals}
\begin{tabular}{lcccc}
\toprule
\textbf{Parameter} & $\hat{\beta}$ & std.~error & $z$-stat & $p$-value \\
\midrule
Openness    & 
$7.625$   & 
$4.42\times10^{-1}$ & 
$17.244$ & 
$0$ \\
Conscientiousness   & 
$-0.188$              & $7.50\times10^{-1}$ & 
$-0.251$ & 
$8.02\times10^{-1}$ \\
Extraversion        & 
$2.497$              &
$6.06\times10^{-2}$ & 
$4.121$ & 
$3.77\times10^{-5}$\\
Neuroticism         & 
$-0.970$              &
$4.19\times10^{-2}$ & 
$-2.312$ & 
$2.08\times10^{-2}$\\
Agreeableness       & 
$1.626$              &
$4.40\times10^{-2}$ & 
$3.698$ & 
$2.17\times10^{-4}$\\
Event Severity      & 
$-1.224$                 &
$8.83\times10^{-1}$ &
$-1.386$ &
$1.66\times10^{-1}$\\
Intercept           &
$-3.072$                 &
$7.46\times10^{-1}$ &
$-4.121$&
$3.78\times10^{-5}$\\
\bottomrule
\end{tabular}
\end{table}

\begin{figure}[!htb]
    \centering
    \includegraphics[width=0.5\linewidth]{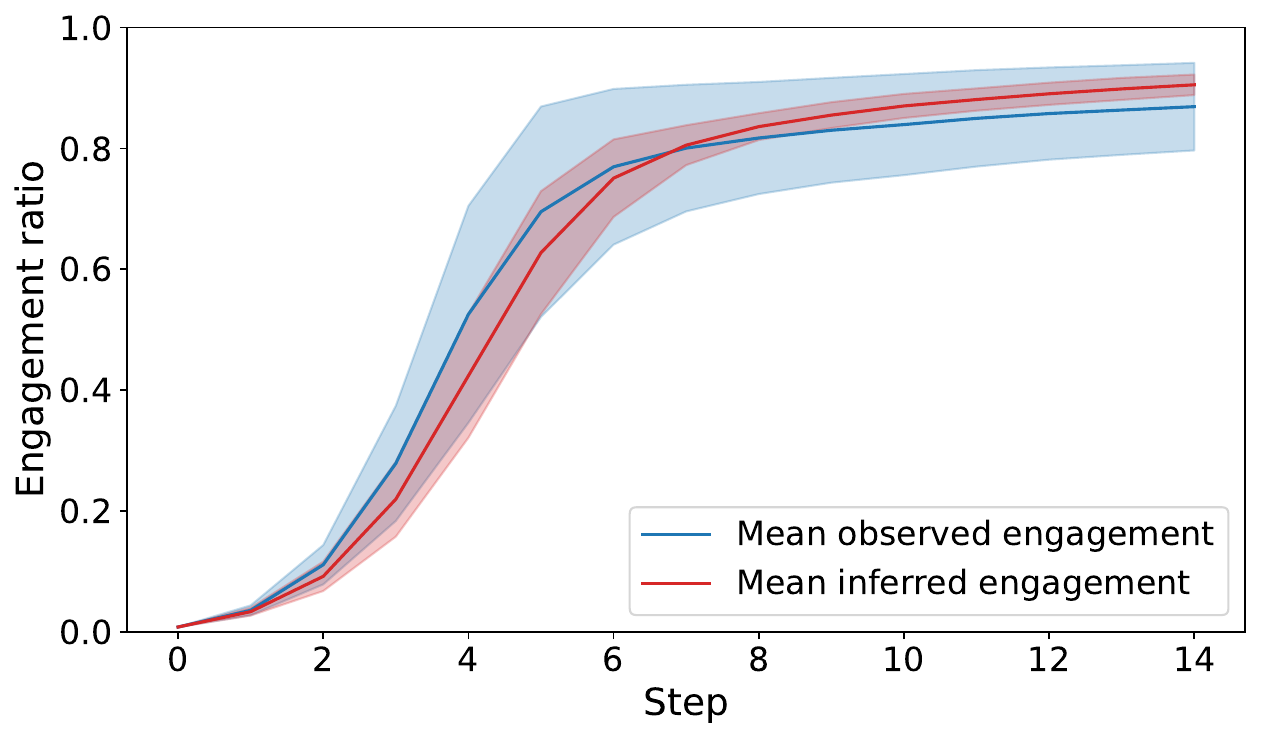}
    \caption{Mean engagement ratio over time across the population from the full logistic model \eqref{model:logistic_32}. The shaded envelope represents $\pm 1$ standard deviation. The inferred results are based on $100$ runs. The inferred curve largely captures the true dynamics, with a slight underestimation in the early stage and a small overshoot at the end.
}
    \label{fig:Logistic_Fit_Population_32Traits}
\end{figure}

Figure \ref{fig:Logistic_Fit_Population_32Traits} shows population-level engagement over time, comparing simulated LLM data (blue) with engagement predicted by the full logistic model (red). To quantitatively assess the goodness of fit, we report the root mean squared error (RMSE) between the observed data and the inferred dynamics. Here, the RMSE is $0.0414$. Overall, the model captures the dynamics well, though it slightly underestimates engagement early on and overestimates it later. 

\begin{figure}[!htb]
    \centering
    \begin{subfigure}[t]{0.48\textwidth}
        \centering
        \includegraphics[width=\linewidth]{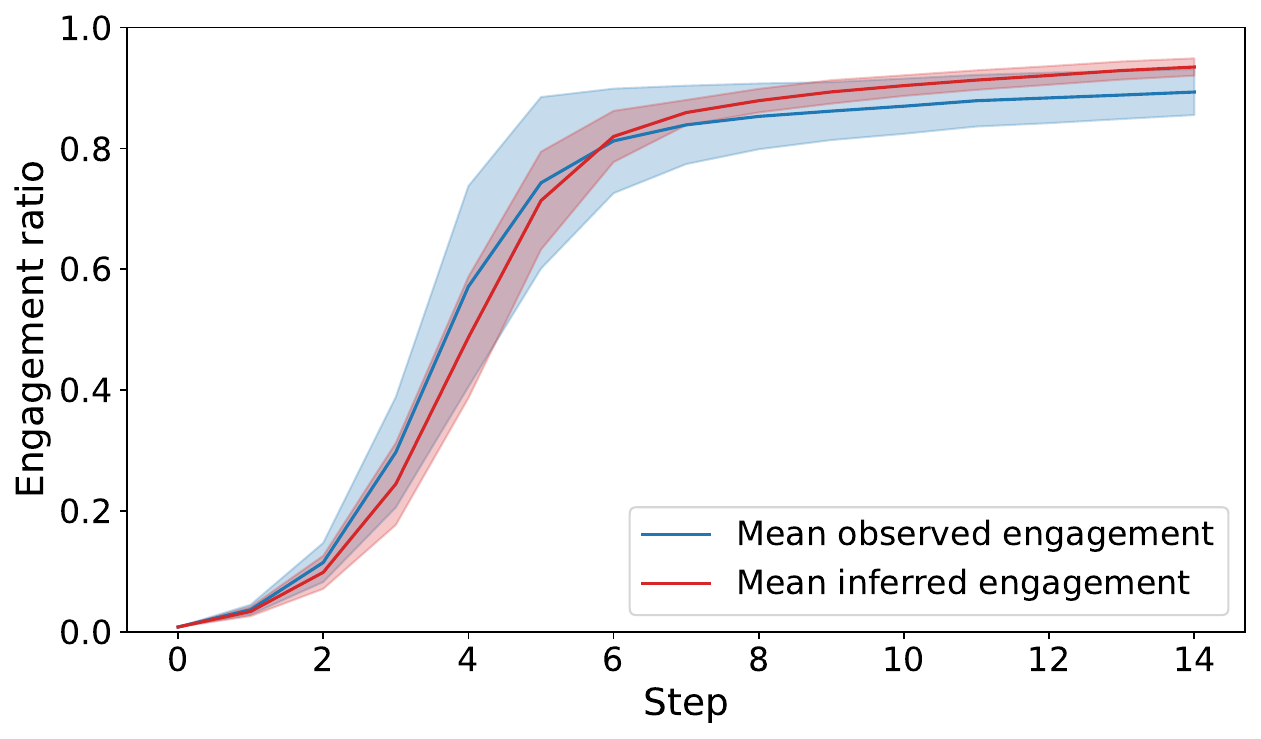}
        \caption{Peaceful event}
        \label{subfig:Logistic_Fit_PeacefulEvent_32Traits}
    \end{subfigure}\hfill
    \begin{subfigure}[t]{0.48\textwidth}
        \centering
        \includegraphics[width=\linewidth]{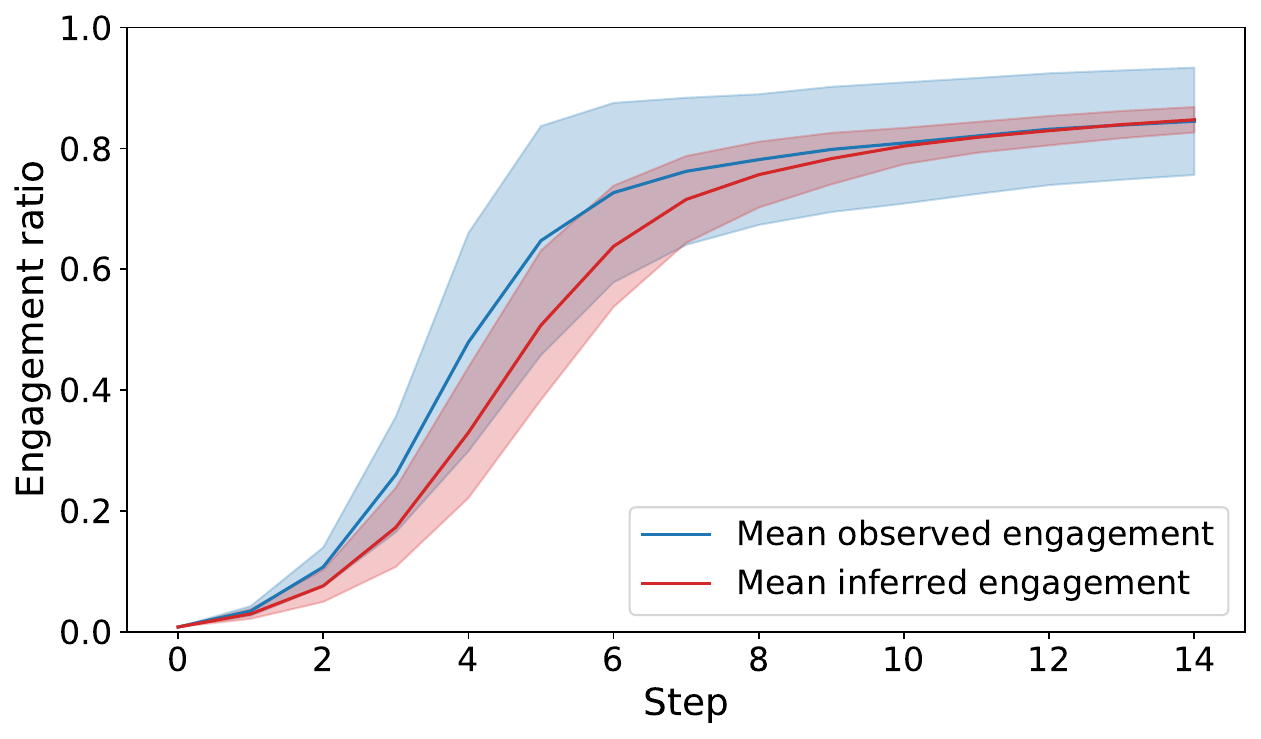}
        \caption{Severe event}
        \label{subfig:Logistic_Fit_SevereEvent_32Traits}
    \end{subfigure}
    \caption{Mean engagement ratio over time based on event type from the full logistic model \eqref{model:logistic_32}. Results shown over $100$ inferred runs, with the shaded envelope representing $\pm 1$ standard deviation. The model predicts better on peaceful events, particularly in the initial period.}
    \label{fig:Logistic_Fit_by_Event_32Traits}
\end{figure}

We also examine the engagement curves disaggregated by event severity. Figure \ref{fig:Logistic_Fit_by_Event_32Traits} shows the predictions by logistic model \eqref{model:logistic_32} alongside the LLM data for each severity group. The RMSE for peaceful event is $0.0367$, whereas that for severe event is $0.0641$. The model matches the dynamics of peaceful events more closely than those of more severe events, especially the initial engagement ratio in the first few time steps.


\subsection{Mean-field model results}
\begin{figure}[!htb]
    \centering
    \includegraphics[width=0.5\linewidth]{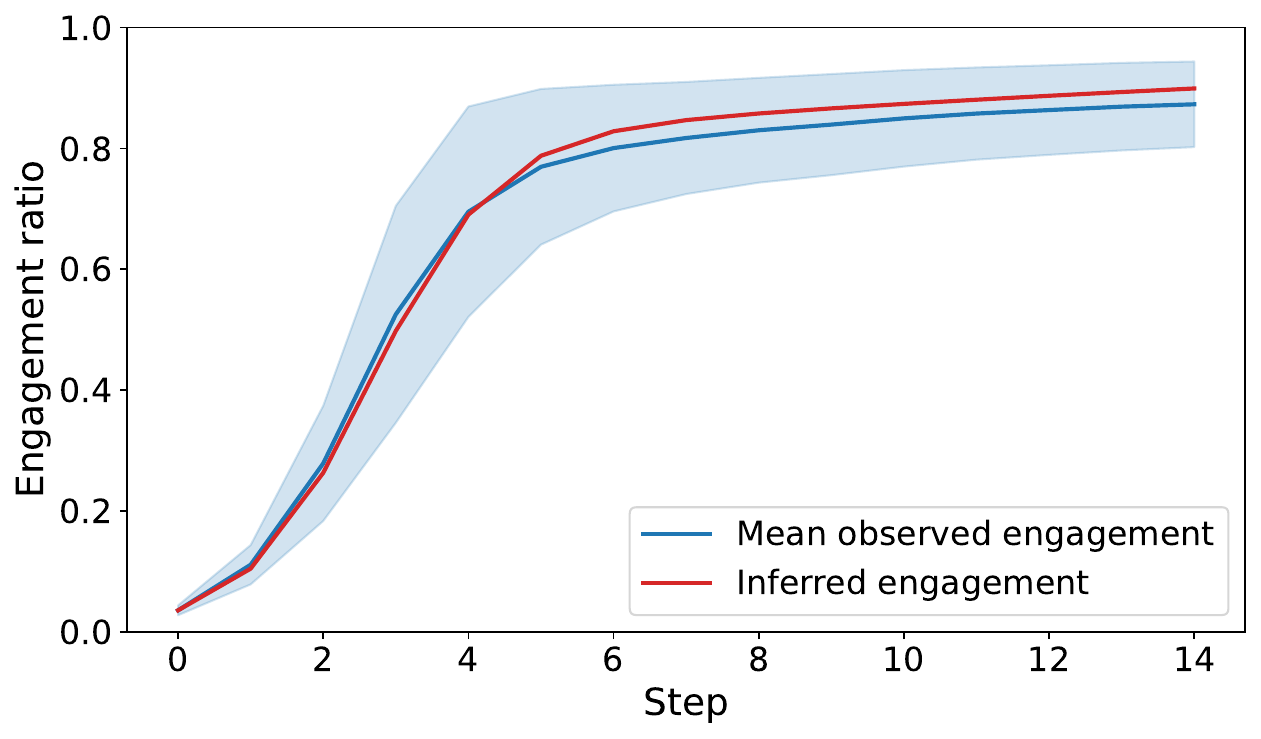}
    \caption{Mean engagement ratio over time at population level as inferred by the mean-field model \eqref{model:MF}. The mean-field model matches the population-level engagement data better than the full logistic model, suggesting a more efficient modeling choice.}
    \label{fig:ODE_Fit_Population}
\end{figure}

We next analyze the mean-field model \eqref{model:MF} fit to the LLM simulation data. Figure \ref{fig:ODE_Fit_Population} shows how engagement evolves over time at the population level, comparing the observed data with the model prediction. The RMSE is $0.0223$. Notably, this two-parameter mean-field model provides a better approximation than the logistic model as shown in Figure \ref{fig:Logistic_Fit_Population_32Traits} (which has RMSE $0.0414$), despite being much simpler. This is in part due to the way the model parameters are estimated. For the mean-field model, the parameters are chosen that directly minimize the square error between the model and data engagement curves (whereas in the logistic model, the parameters are chosen to minimize the likelihood of the individual-level engagement probability).

\begin{figure}[!htbp]
    \centering
    \begin{subfigure}[t]{0.48\textwidth}
        \centering
        \includegraphics[width=\linewidth]{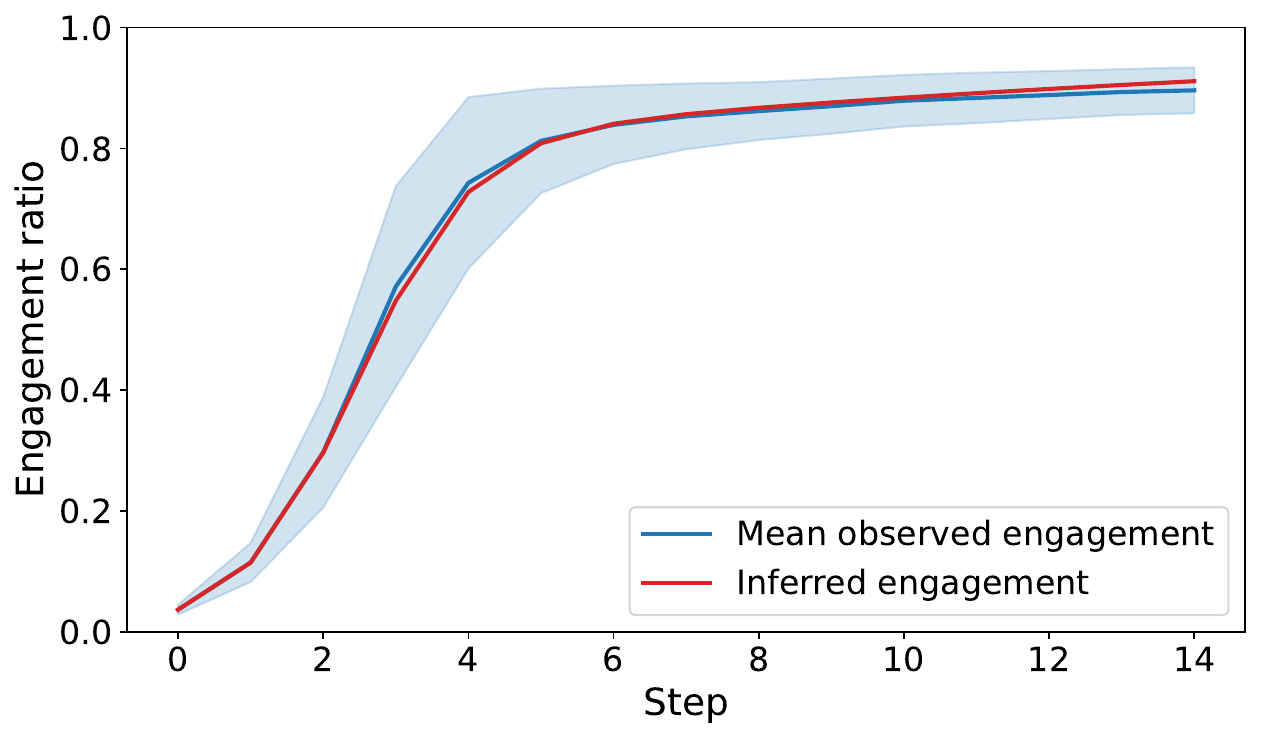}
        \caption{Peaceful event}
        \label{subfig:ODE_Fit_EventType1}
    \end{subfigure}\hfill
    \begin{subfigure}[t]{0.48\textwidth}
        \centering
        \includegraphics[width=\linewidth]{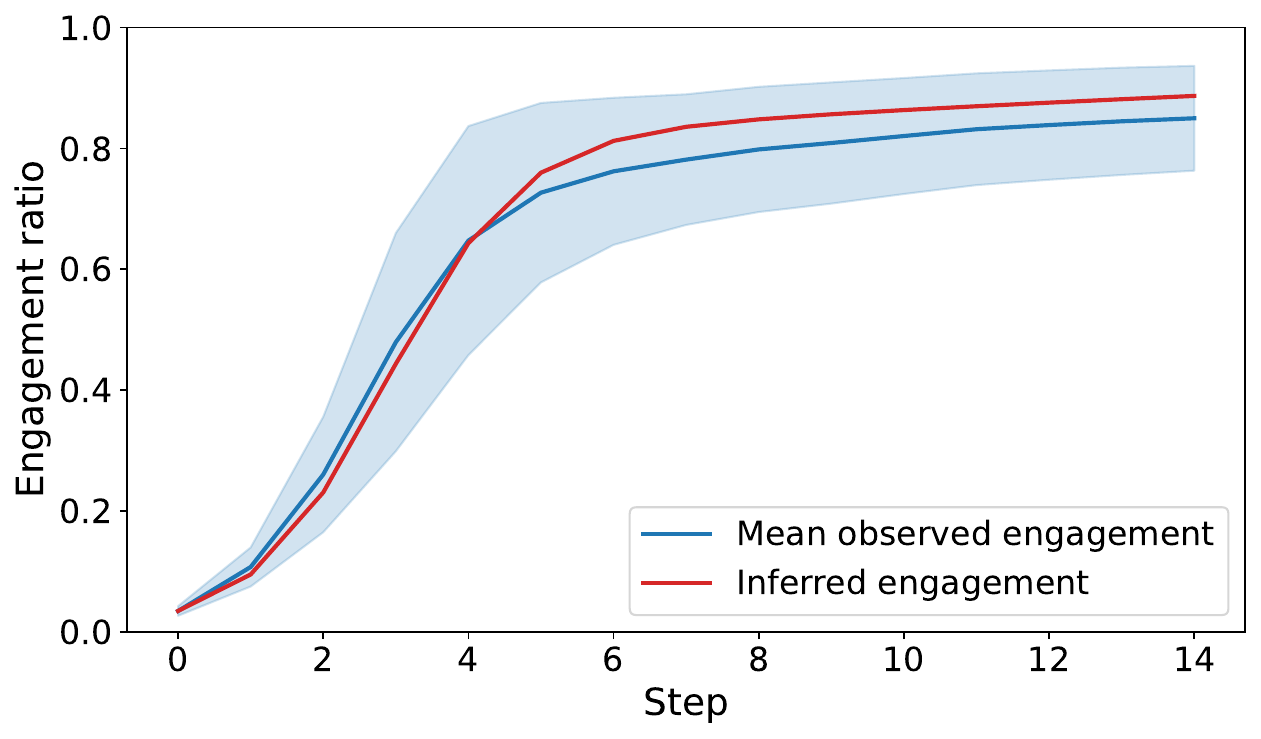}
        \caption{Severe event}
        \label{subfig:ODE_Fit_EventType2}
    \end{subfigure}
    \caption{Mean engagement ratio over time based on event type as inferred by the mean-field model \eqref{model:MF}. The model's inferred dynamics match better with the data for peaceful events.}
    \label{fig:ODE_Fit_EventTypes}
\end{figure}

Once again, we break down the engagement dynamics by the severity of the news event. Figure \ref{fig:ODE_Fit_EventTypes} compares the model’s inferred trajectories with the observed data at each severity level. The RMSE for peaceful event is $0.0098$ and for severe event is $0.0374$. For this model, we see that the inference based on event type is also more accurate than that provided by the logistic model \eqref{model:logistic_32}, whose results are shown in Figure \ref{fig:Logistic_Fit_by_Event_32Traits} with RMSE $0.0367$ and $0.0641$, respectively. We also observe that the model's approximation aligns especially well with peaceful events. In fact, in this case, the predicted dynamics match the observed behavior almost exactly, whereas the fit becomes less accurate as event severity increases.


\begin{figure}[!htbp]
    \centering
    \begin{subfigure}[t]{0.48\textwidth}
        \centering
        \includegraphics[width=\linewidth]{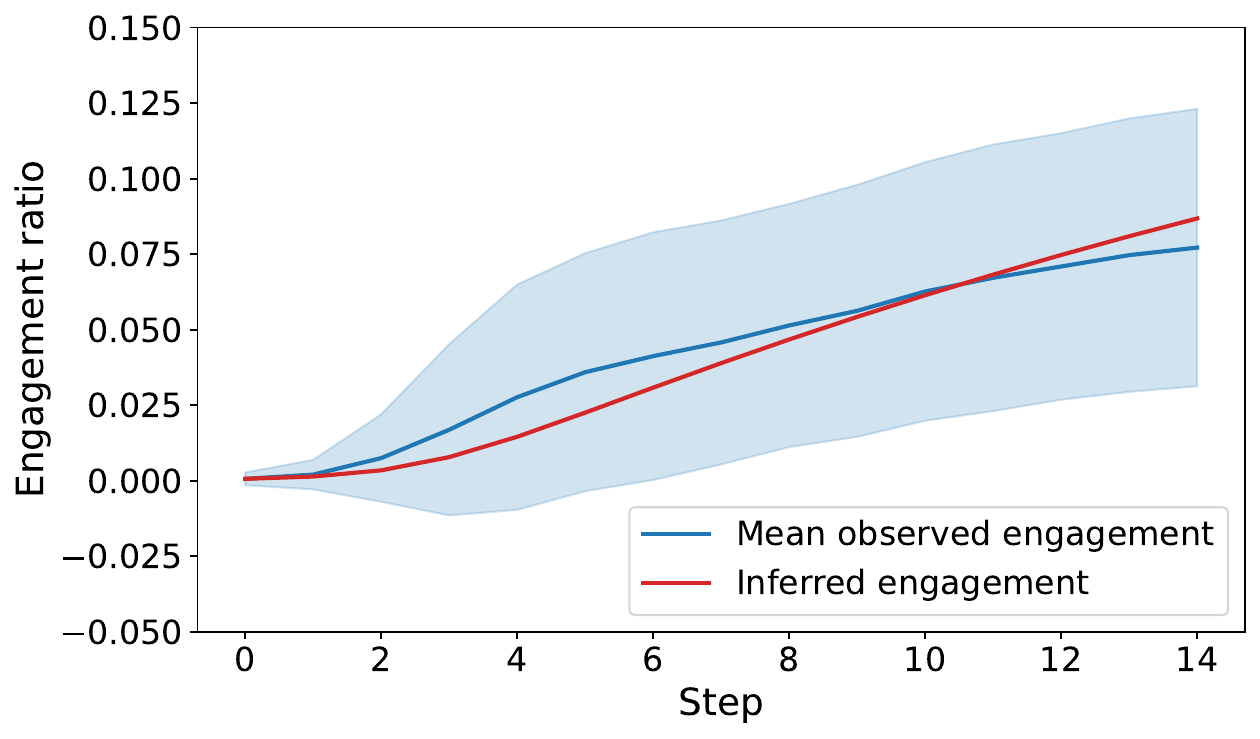}
        \caption{Trait group 1}
        \label{subfig:ODE_Fit_TraitGroup1}
    \end{subfigure}\hfill
    \begin{subfigure}[t]{0.48\textwidth}
        \centering
        \includegraphics[width=\linewidth]{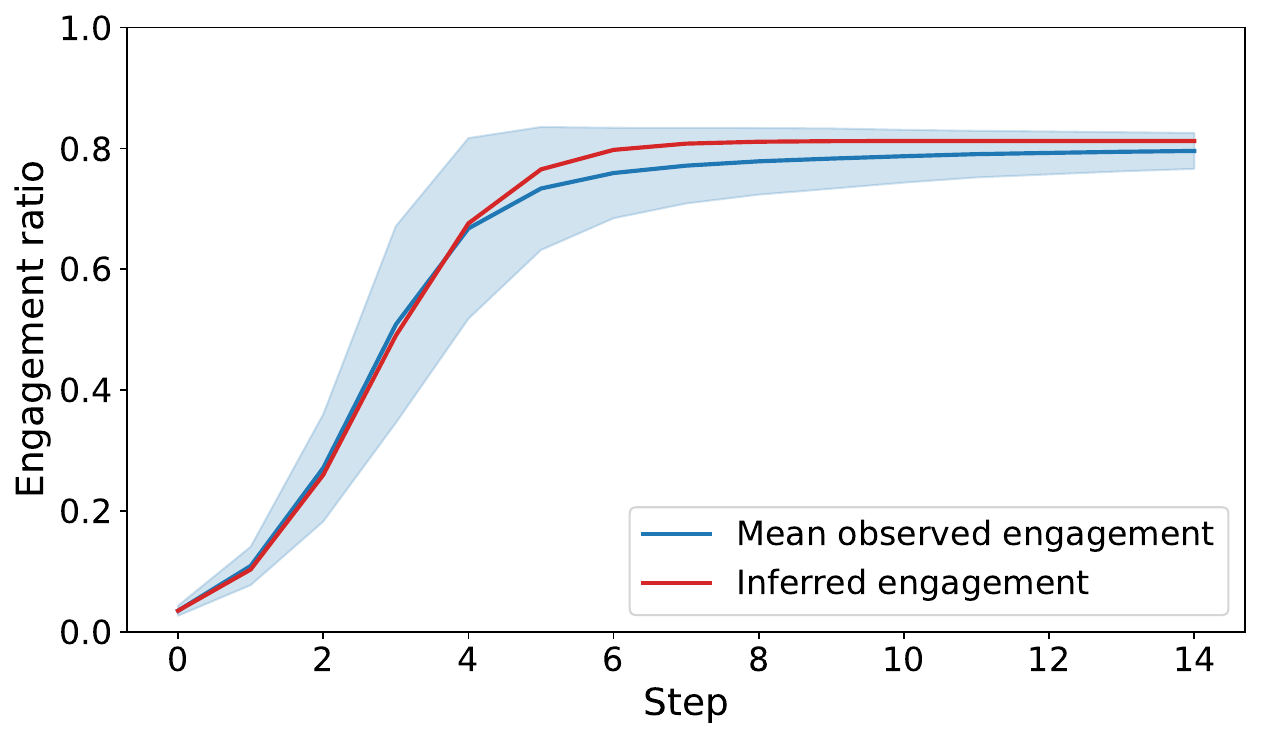}
        \caption{Trait group 2}
        \label{subfig:ODE_Fit_TraitGroup2}
    \end{subfigure}
    \caption{Mean engagement ratio over time based on trait groups as inferred by the mean-field model \eqref{model:MF}. Trait group $2$ has a much higher transmission rate than group $1$, resulting in a much higher engagement ratio over time.}
    \label{fig:ODE_Fit_TraitGroups}
\end{figure}

For the mean-field model, we also examine how engagement evolves within the two trait groups. Figure \ref{fig:ODE_Fit_TraitGroups} compares the inferred dynamics with the observed data disaggregated by trait groups. The results show a clear separation between the groups: agents in group $1$ consistently have much lower engagement ratios, while agents in group $2$ are noticeably more active in sharing information with their neighbors. As a complement, we present in Table \ref{tab:params_MF} the values of the two transmission parameters inferred for the mean-field model using the training data, where we can clearly see that group $1$ has a low transmission rate, while the rate for group $2$ is much higher. Since the engagement levels for the two groups have different scales, we report in addition to RMSE the normalized root mean square error (NRMSE), which is calculated as the RMSE normalized by the mean observed value. For trait group $1$, the RMSE is $0.0072$ and the NRMSE is $0.1696$; for trait group $2$, the RMSE is $0.0236$ and the NRMSE is $0.0377$. From these results, we see that while group $1$ shows better agreement in the absolute terms, its relative discrepancy is larger.


\begin{table}[!htbp]
\centering
\caption{Inferred parameter values for the mean-field model.}
\label{tab:params_MF}
\begin{tabular}{lcc}
\toprule
 & $\tau_1$ & $\tau_2$ \\
\midrule
Inferred value & $0.01269$ & $0.28520$ \\
\bottomrule
\end{tabular}
\end{table}

The separation between these two groups helps explain the global dynamics observed in Figure \ref{fig:ODE_Fit_Population}. While stability analysis indicates that all individuals eventually become engaged, Figure \ref{fig:ODE_Fit_Population} shows a flattening of the curve at an engagement ratio between $0.8$ and $0.9$. The initial rapid increase in engagement is driven by group $2$, who quickly become engaged. However, group $1$ individuals become engaged more slowly, as reflected by the gradual increase in the global engagement curve after the first few time steps.

\section{Conclusion}\label{sec:conclusion}
In this work, we studied the propagation of news shared on a multi-agent network of LLMs.  We varied the LLM personalities defined in the prompt, network connections, and the type of news shared, resulting in different patterns of overall engagement. We then fit these engagement patterns with an agent-based logistic model and a mean-field differential equation model. We found that these models capture the main features of how information spreads, with the two-parameter mean-field model providing the closest match to the simulation data. We also found clear differences in how agents behave: peaceful events are shared more easily than violent ones, and open and extraverted personality groups are much more likely to spread information than others. The overall engagement growth is driven initially by the rapid engagement of the open group, followed by a slow increase in engagement by the remaining LLMs in the network.

These results suggest that large groups of LLM agents may exhibit stable, predictable spreading patterns, and that low-dimensional mathematical models can help us understand and even forecast how information propagates through LLM networks. Although we focused on \textit{OpenAI}'s \textit{gpt-3.5-turbo-1106} model in this study, we believe that our modeling framework is well-suited for studying a broad range of LLMs (and this is a good direction for future research).  Mean-field models have the potential for increasing our understanding of complex information cascades in real online networks, and exploring how human and AI agents interact at large scales.

While this work focused on a network consisting of only LLMs, one promising direction for future research is to incorporate interactions between LLMs and human agents. In one approach, LLMs can be trained on human-generated social media data, and the behaviors of groups of LLMs can represent key aspects of human collective behaviors in simulations of information propagation across online networks \cite{AI_collective_behavior}. An alternative approach would be to conduct controlled experiments with real human participants and a network experiment platform such as oTree \cite{priniski2024online,jha2025simulating}.  This type of multi-agent LLM network simulation would provide a controllable testbed for studying the propagation of information in complex, online networks where humans and LLMs interact.

\vspace{20pt}
\noindent \textbf{Data accessibility.} 
Code for reproducing the results in this article is available at: https://github.com/Moyi-Tian/Learning-LLM-Dynamics

\vspace{5pt}
\noindent \textbf{Acknowledgements.}  
This work was supported by the U.S. National Science Foundation Division of Mathematical Sciences [grant number 2042413] and the Air Force Office of Scientific Research Multidisciplinary University Research Initiative [grant number FA9550-22-1-0380].

\appendix

\section{Logistic model based on trait-profile groups} \label{app:logistic_trait_group_based}
Here we consider a two-group version of Equation \eqref{model:logistic_32}, analogous to the two-group mean-field model. We model the probability that an ``unengaged" node becomes ``engaged" upon exposure to an engaged neighbor using a trait-profile-group-dependent logistic model.

Using the same notation as before, we introduce $\beta_{\mathbf{g}}$ as the learned weight specific to trait profile group $\mathbf{g}$. Assuming two trait groups $g_1$ and $g_2$, the probability of engagement given traits $\mathbf{t}$ and event severity $s$ is modeled as
\begin{align} \label{model:logistic_2}
    \Pr(y = 1 \mid \mathbf{t}, s) = \sigma\!\left(\mathbf{1}_{\mathbf{t} \in g_1} \cdot \beta_{\mathbf{g_1}} + \mathbf{1}_{\mathbf{t} \in g_2} \cdot \beta_{\mathbf{g_2}} + \beta_s \cdot s\right),
\end{align}
where $\sigma(z) = \frac{1}{1 + e^{-z}}$ is the logistic sigmoid function.

Given $N$ datapoints $\{(\mathbf{t}_i, s_i, y_i)\}_{i=1}^N$, the parameters are estimated by minimizing the negative log-likelihood
\[
\mathscr{L} = - \sum_{i=1}^N \left[ y_i \log p_i + (1 - y_i) \log(1 - p_i) \right],
\]
where $p_i = \Pr(y_i \mid \mathbf{t}_i, s_i)$ is determined by Equation \eqref{model:logistic_2}.  The parameters to be learned are $\beta_{\mathbf{g_1}}, \beta_{\mathbf{g_2}} \in \mathbb{R}$ (one for each personality profile group), and $\beta_s \in \mathbb{R}$ (severity effect).

\begin{figure}[!htbp]
    \centering
    \includegraphics[width=0.49\linewidth]{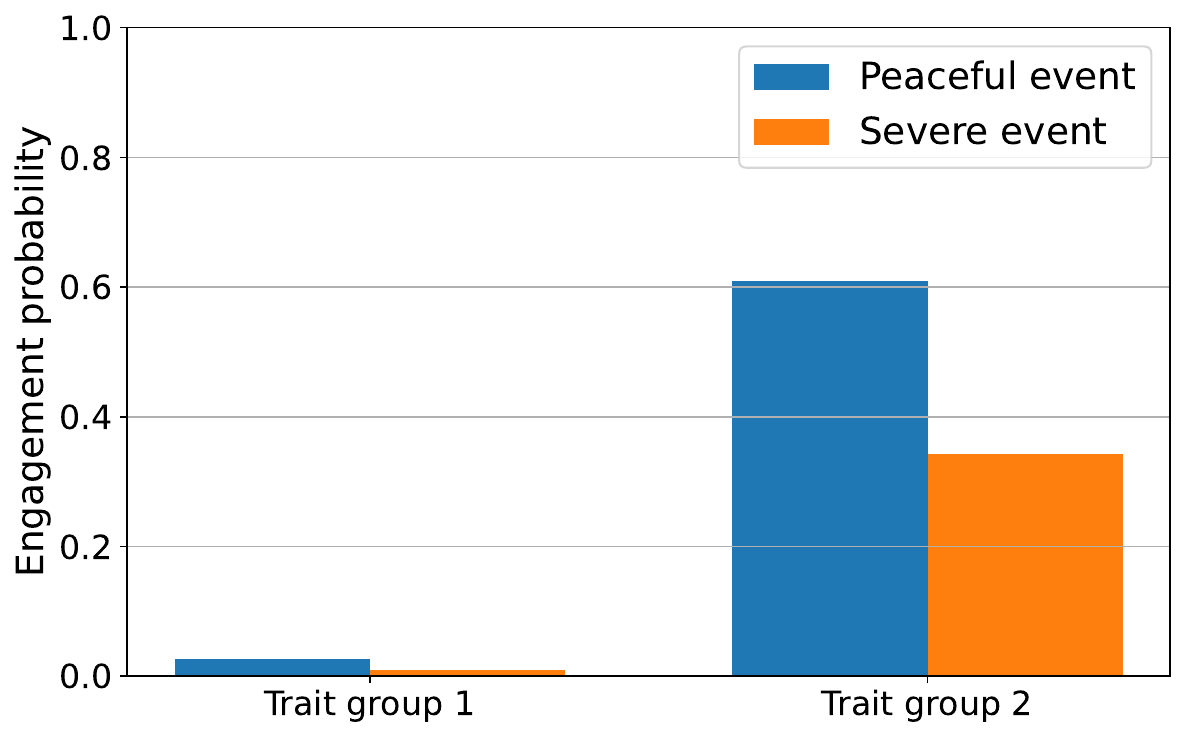}
    \caption{Learned probability by trait group and event type from the simplified logistic model \eqref{model:logistic_2}. Individuals in trait group $1$ have a much lower probability of becoming engaged, meaning starting to actively spread the news with neighbors, compared with those in group $2$.}
    \label{fig:Logistic_Fit_Probability_2TraitGroups}
\end{figure}

In Figure \ref{fig:Logistic_Fit_Probability_2TraitGroups}, we show the learned probabilities for the two trait groups, disaggregated by event severity.  Here we observe a similar separation between group $1$ and $2$ as we did for the mean-field model, where group $1$ has a much lower probability for sharing news.

\begin{figure}[!htb]
    \centering
    \includegraphics[width=0.48\linewidth]{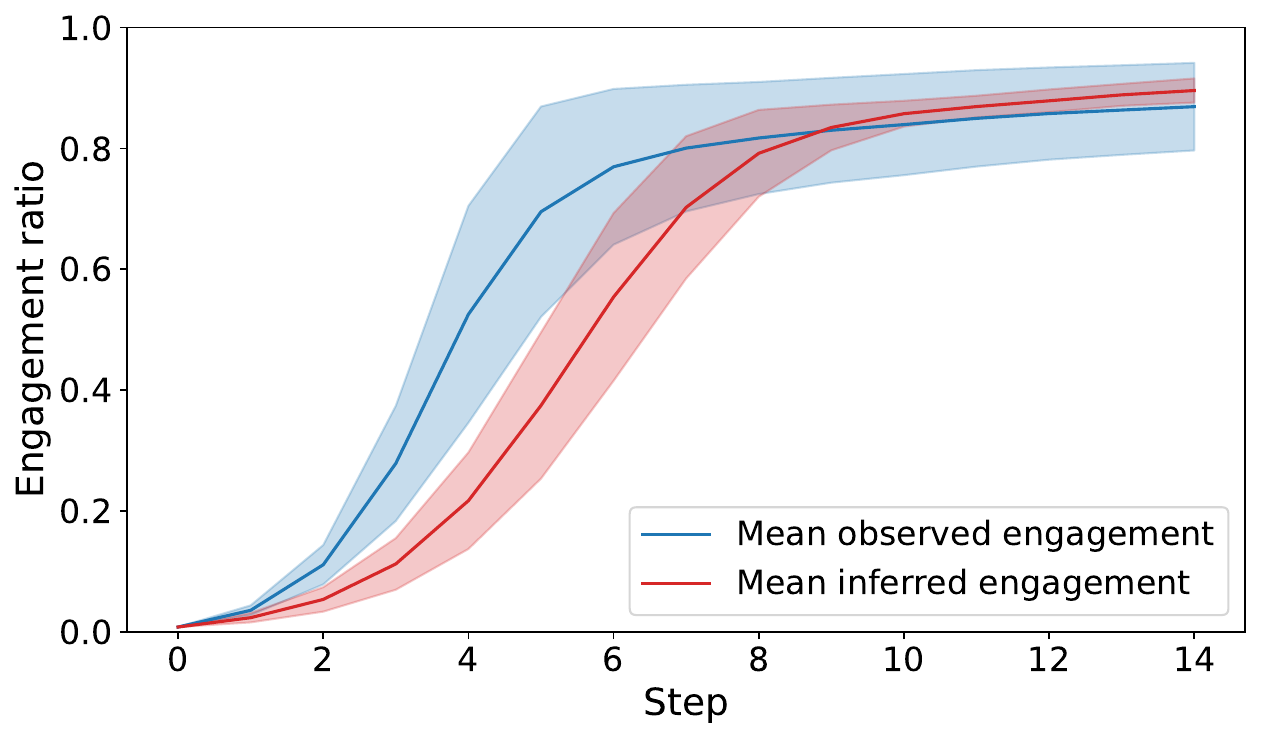}
    \caption{Mean engagement ratio over time at the population level from the simplified logistic model \eqref{model:logistic_2}, with $100$ inferred runs.}
    \label{fig:Logistic_Fit_Population}
\end{figure}

\begin{figure}[!htb]
    \centering
    \begin{subfigure}[t]{0.46\textwidth}
        \centering
        \includegraphics[width=\linewidth]{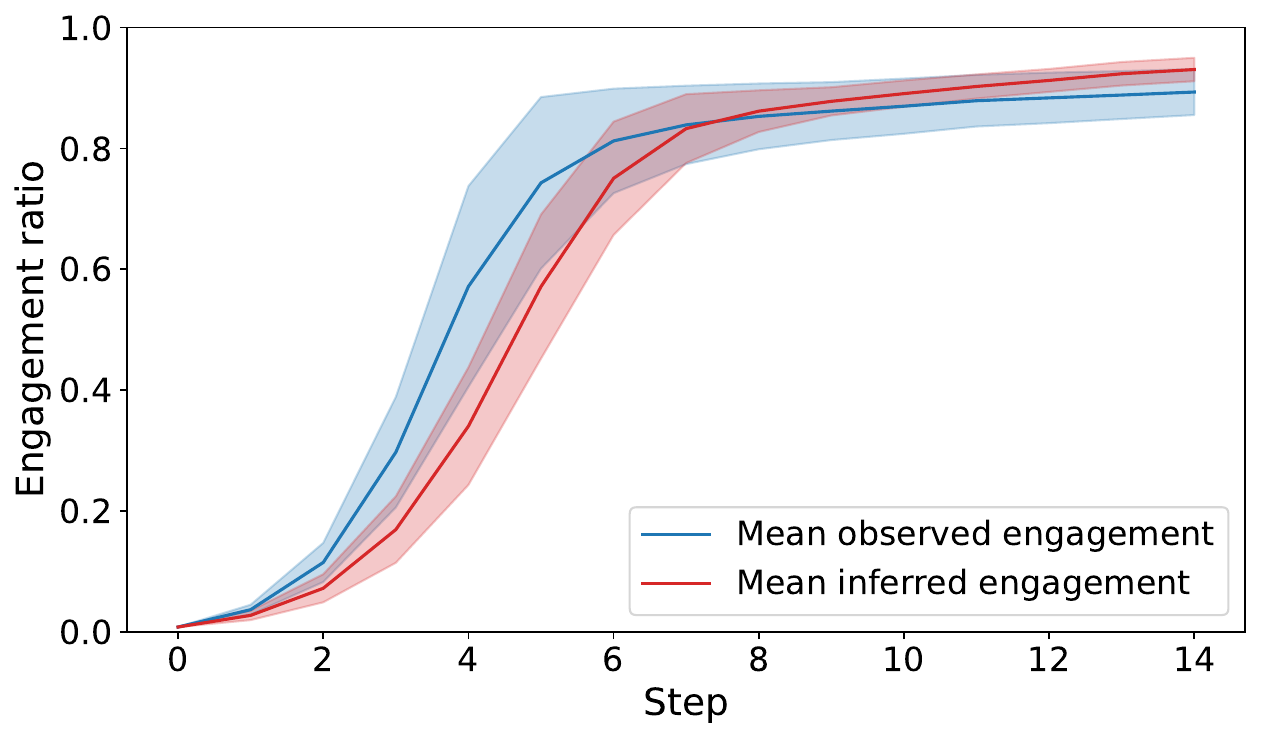}
        \caption{Peaceful event}
        \label{subfig:Logistic_Fit_EventType1}
    \end{subfigure}\hfill
    \begin{subfigure}[t]{0.46\textwidth}
        \centering
        \includegraphics[width=\linewidth]{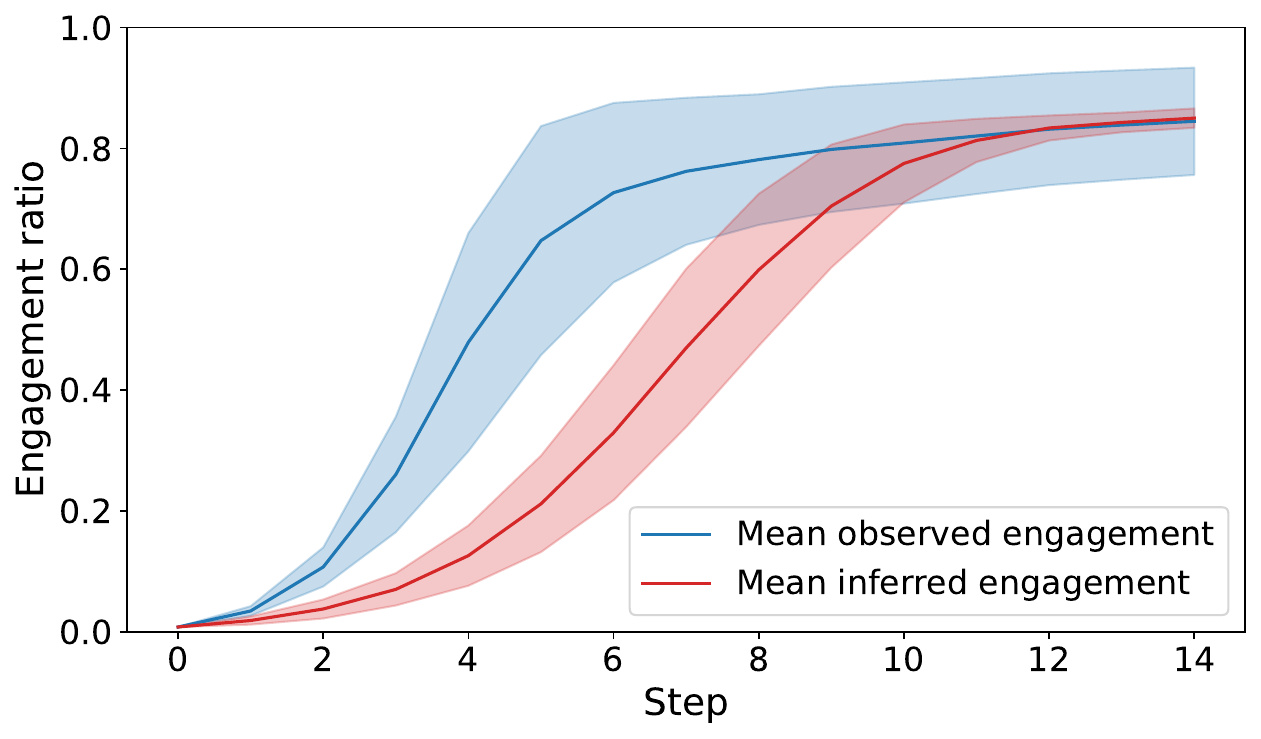}
        \caption{Severe event}
        \label{subfig:Logistic_Fit_EventType2}
    \end{subfigure}
    \caption{Mean engagement ratio over time based on event type from the simplified logistic model \eqref{model:logistic_2}, with $100$ inferred runs.}
    \label{fig:Logistic_Fit_EventTypes}
\end{figure}

In Figure \ref{fig:Logistic_Fit_Population}, we present the population-level engagement dynamics of the fitted model compared to the observed data. The RMSE is $0.1387$. In Figure \ref{fig:Logistic_Fit_EventTypes}, we plot the engagement curves disaggregated by severity levels. The RMSE is $0.0856$ for peaceful event and $0.2070$ for severe event. Here we observe that the fit of the model is worse than that of the $32$-parameter logistic model, as well as the $2$-parameter mean-field model.  As mentioned above, this is due to the way the model is estimated; the parameters are chosen to minimize the likelihood of the individual-level engagement probability, rather than the mean-square error of the engagement curves (and for the logistic model, $2$ parameters are not enough to accurately capture the global dynamics).

\begin{figure}[!htb]
    \centering
    \begin{subfigure}[t]{0.46\textwidth}
        \centering
        \includegraphics[width=\linewidth]{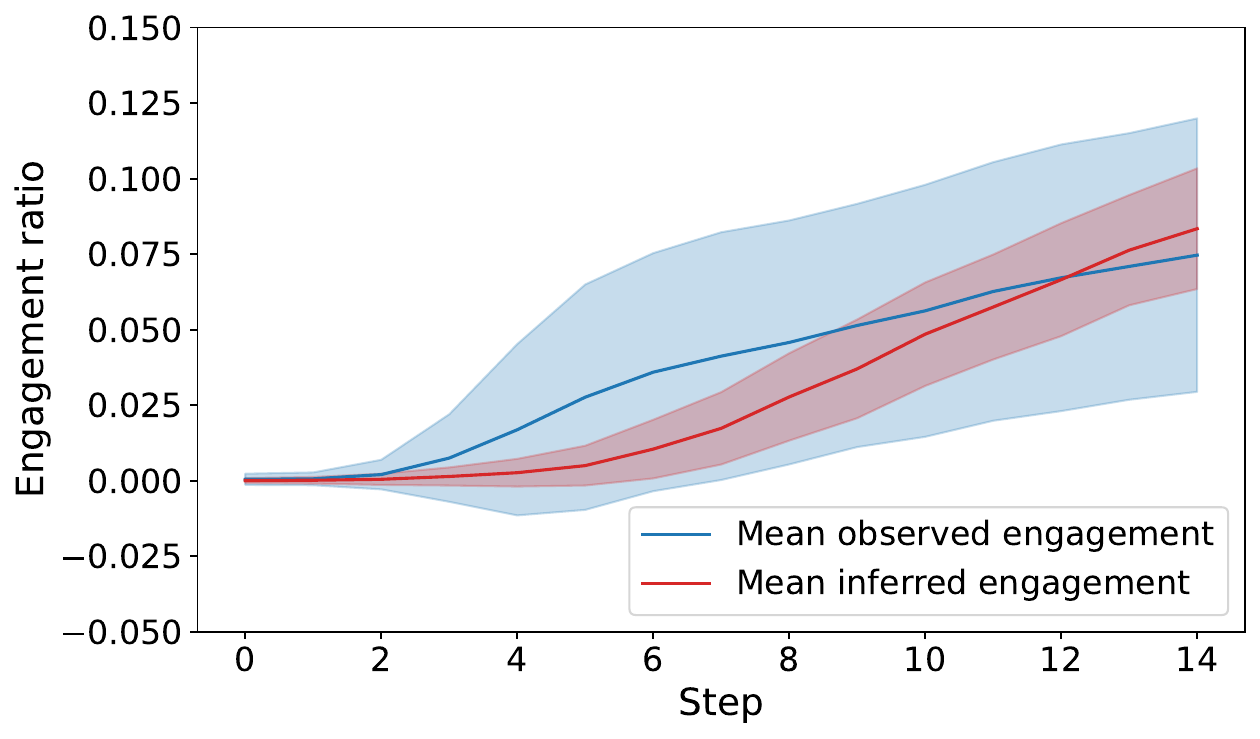}
        \caption{Trait group 1}
        \label{subfig:Logistic_Fit_TraitGroup1}
    \end{subfigure}\hfill
    \begin{subfigure}[t]{0.46\textwidth}
        \centering
        \includegraphics[width=\linewidth]{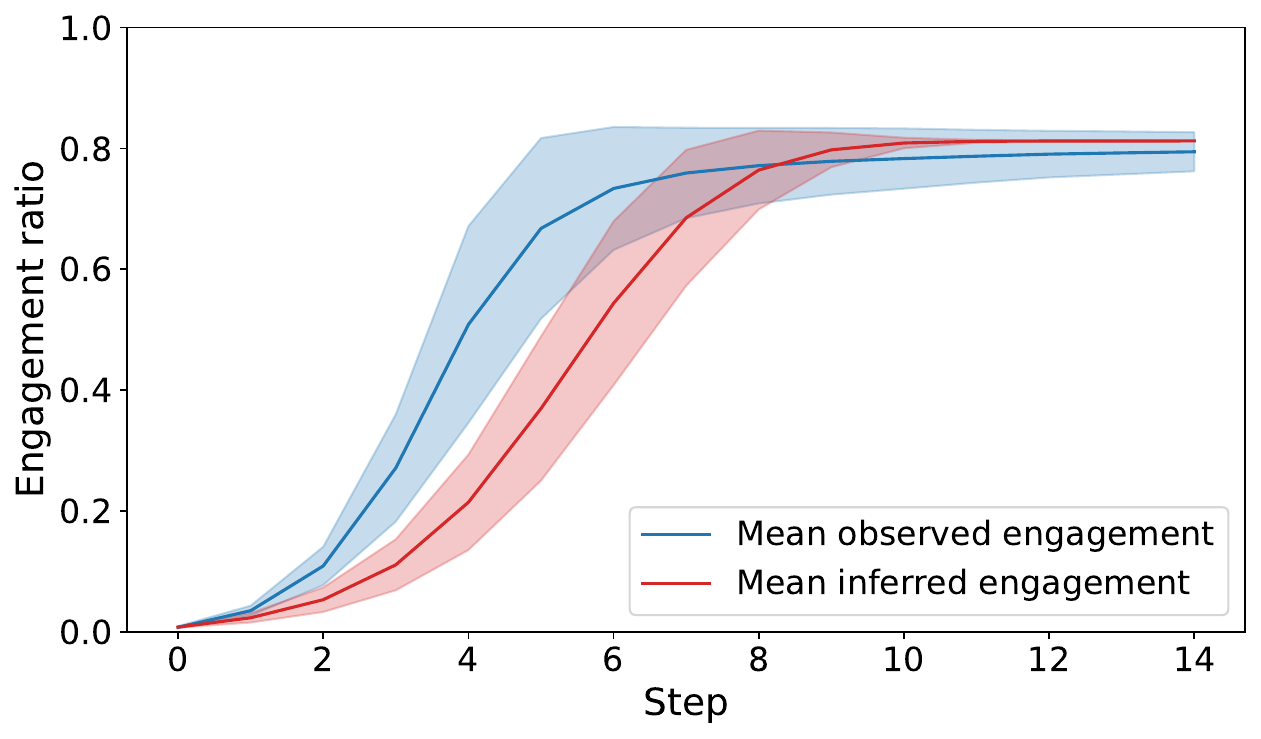}
        \caption{Trait group 2}
        \label{subfig:Logistic_Fit_TraitGroup2}
    \end{subfigure}
    \caption{Mean engagement ratio over time based on trait groups from the simplified logistic model \eqref{model:logistic_2}, with $100$ inferred runs.}
    \label{fig:Logistic_Fit_TraitGroups}
\end{figure}

In Figure \ref{fig:Logistic_Fit_TraitGroups}, we present observed and the model-inferred engagement dynamics disaggregated by the $2$ trait groups. For group $1$, the RMSE is $0.0134$ and NRMSE is $0.3586$; for group $2$, the RMSE is $0.1288$ and NRMSE is $0.2250$. Here, we again observe a rapid increase in engagement for group $2$, followed by a slower increase in engagement for group $1$ (though the fit is not as good as in the mean-field model).


\FloatBarrier
\bibliographystyle{plain}
\bibliography{refs}

\end{document}